\documentstyle[12pt]{article}
\title{On genus-one-corrected extremal black holes
and the correspondence principle}
\author {Mikhail Z. Iofa\thanks{E-mail address: iofa@theory.npi.msu.su}\\
Nuclear Physics Institute \\
Moscow State University\\
Moscow 119899, Russia \\
 and \\
Leopoldo A. Pando Zayas\thanks{E-mail address: leopoldo@grg1.phys.msu.su}\\
Department of Theoretical Physics\\
Moscow State University\\
Moscow 119899, Russia}
\date{}

\def\pd{\partial_{\mu}}
\def\pa{\partial}

\def\p{\phi}

\def\l{\lambda}
\def\s{\sigma}

\def\g{\gamma}

\def\D{\Delta}
\def\a{\alpha}
\def\b{\beta}

\begin{document}
\maketitle

\begin{abstract}
We discuss charged black-hole solutions to the equations of motion of the
string-loop-corrected effective action. At the string-tree level, these
solutions provide backgrounds for the "chiral null model". The effective
action contains gravity, dilaton and moduli fields. Analytic solutions   of
the one-loop-corrected equations of motion are presented for the
extremal magnetic and dyonic black holes.   Using the fact that
 in magnetic solution the loop-corrected dilaton 
is non-singular at the origin, 
we apply the correspondence  principle to show that the entropy of the
loop-corrected magnetic black hole can be interpreted as the microscopic
entropy of the D-brane system.

\end{abstract}


\vfill
NPI-MSU-97-38/489\\
hep-th/9711086

\pagebreak
\section{Introduction}
At present, string theory is considered  the best candidate for
 a fundamental theory that would provide a consistent quantum theory of
gravity unified with other interactions \cite{gsw}. Thus, there is a problem of 
understanding how the intrinsically stringy effects modify  Einstein
gravity.
In this paper we are going to focus on two of these effects: the presence of
scalar fields such as the dilaton and the moduli; and higher-genus 
contributions modifying  the tree-level effective action. 
 We  consider a class of black-hole solutions obtained in the "chiral null
models" embedded into the (compactified) heterotic string theory \cite{bh}
which includes the  majority of all superstring
black holes, in particular, electric, magnetic and dyonic black holes.
Within the "chiral null models" it was possible to show that the
Bekenstein-Hawking entropy can be interpreted as statistical entropy
of string oscillating states \cite{sus} and \cite{lw} 
(for another class of models see review \cite{ho} 
and references therein). 

In paper \cite{cvts1} a class of solutions of the model effective action
for the abelian field interacting with the scalar moduli fields in flat 4D
space-time was investigated. Moduli-dependence of gauge couplings mimicked
 higher-loop string effects. 
In paper
\cite{chan}, using a model string effective action 
which includes a small constant threshold correction to the gauge coupling, 
an electric black hole solution was considered. However, no modulus 
dependence of the model effective action was taken into 
account and
the coupling of the dilaton to the electric field strength was assumed to
be small. 

In this paper we investigate a class of "realistic" charged black-hole 
solutions 
obtained by  using  a  4D effective action  which includes
dilaton and modulus fields. The gauge fields enter the action  with 
gauge couplings that include threshold corrections 
 modifications modeling  string-loop effects.

 In section 2 we review the structure of the 4D dyonic black hole
in heterotic string theory provided by the chiral null model. In section
3 we consider the structure of the genus-one (torus topology) correction
to the low-energy string effective action and discuss its dependence on the
moduli.

 Section 4 is devoted to calculation of  modifications of geometric and
thermodynamic properties of the black-hole solutions to the
string-loop-corrected equations of motion.

 As compared to previous results, the present  investigation is, 
to our knowledge, the first one where string effects are considered for a
 realistic class of superstring black hole solutions.

Using the explicit loop-corrected solution for the extremal magnetic black hole,
we show that dilaton is non-singular at the origin. This opens a possibility
to apply to this system the correspondence principle and to show that the
geometrical entropy of magnetic black hole is reproduced as the microscopic
entropy of the weakly-interacting D-brane system.
\section{Dyonic 4D black hole in toroidally compactified heterotic string
theory}

In this paper we discuss  black hole  solutions obtained in
conformal  chiral null models with curved transverse part. These can be 
considered  embedded in  heterotic string theories.  The 4D low energy
effective action is 
obtained by dimensional reduction of the  10D effective action of heterotic
string theory compactified on $T^6=T^2\times T^4$ (see \cite{bh} for 
details) \footnote{We discuss a simplified case without 
the axion field which can be  also included in the chiral null model.}.
The 4D low energy effective action (in the Einstein frame) is
\begin{eqnarray}
\label{ea}
S&=&\int d^4 x\sqrt{-g}\left(R-\frac{1}{2}(\pd\p)^2-(\pd\s)^2-(\pd\g)^2
\right. \nonumber \\
&-&\frac{1}{4}e^{-\p+2\g}(F_{(1)})^2-\frac{1}{4}e^{-\p+2\s}(F_{(2)})^2
\nonumber \\
&-& \left.  \frac{1}{4}e^{-\p-2\g}(F_{(3)})^2
-\frac{1}{4}e^{-\p-2\s}(F_{(4)})^2\right),
\end{eqnarray}
where $\p$ is the dilaton field, $\s$ and $\g$ are moduli related to the
radii of the $T^2$. Here $F_{(1)}$ and $F_{(3)}$ are  magnetic
fields strengths, and  $F_{(2)}$ and $F_{(4)}$ are electric field strengths.
The nontrivial 6D backgrounds of the chiral null model (the remaining part
of the metric is flat) obtained as solutions of equations
of motion of this action are
\begin{eqnarray}
\label{back}
ds^2&=&-\l(r)dt^2 +\l^{-1}(dr^2+r^2d\Omega_2^2)\nonumber \\
\l^2(r)&=&FK^{-1}kf^{-1}\nonumber \\
2\p&=&\ln FK^{-1}fk^{-1} \nonumber \\
e^{2\s}&=&FK \nonumber \\
e^{2\g}&=&fk
\end{eqnarray}
where $F^{-1}, K, f$ and $k^{-1}$ are harmonic functions satisfying the
equations
$\delta^{ij}\pa_i\pa_j K=0$ for $i,j=1,2,3$, etc..  The gauge
fields electric and magnetic fields  potentials are $A_{(2)}=(K^{-1},0)$ , $A_{(4)}=(F,0) $, and
$A_{(1)}=(0,a_s)$ , $A_{(3)}=(0,b_s)$, $s=1,2,3$.  The latter satisfy
the relations
\begin{eqnarray}
2\pa_{[p}b_{q]}&=&-\epsilon_{pqs}\pa^s f\nonumber \\
2\pa_{[p}a_{q]}&=&-\epsilon_{pqs}\pa^s k^{-1}.
\end{eqnarray}
Electric and magnetic charges are obtained from  the
asymptotics  of the field strengths calculated with the gauge potentials
$A_{(2)}, A_{(4)}$ and $A_{(1)}, A_{(3)}$  respectively.

\section{One-loop corrections}

The tree level effective action in string theory receives string-loop 
corrections
due to integration over  world sheets of higher  topologies \cite{ka}. Here we 
are going to consider  the one-string-loop corrections to the gauge couplings 
\footnote{A more
detailed analysis of relevant threshold corrections to the effective action
will be given elsewhere.}. The effective action is
\begin{eqnarray}
\label{oneloop}
S&=&\int d^4 x\sqrt{-g}\left(R-\frac{1}{2}(\pd\p)^2-(\pd\s)^2-(\pd\g)^2
\right. \nonumber \\
&-&
\frac{1}{4}(e^{-\p+2\g}+\D)(F_{(1)})^2-\frac{1}{4}(e^{-\p+2\s}+\D)(F_{(2)})^2 \nonumber \\
&-& \left.\frac{1}{4}(e^{-\p-2\g}+\D)(F_{(3)})^2-\frac{1}{4}(e^{-\p-2\s}+\D)(F_{(4)})^2\right)
\end{eqnarray}
For toroidal compactification, the one-loop correction
to the action (1) is                        
\begin{equation}
\Delta=-b\ln(T_2|\eta(T)|^4U_2|\eta(U)|^4),
\end{equation}
where $\eta$ is the Dedekind $\eta$-function and the moduli $T$ and $U$ are,
\begin{eqnarray}
T&=&T_1+iT_2=B+i\sqrt{det G}\nonumber \\
U&=&U_1+iU_2=(G_{12}+i\sqrt{det G})/G_{11}.
\end{eqnarray}
Here, $G_{ij}$ are the components of the  "internal"
metric, which in our case are  $G_{12}=0,\, G_{22}=fk=e^{2\g}$ and 
$G_{11}=FK=e^{2\s}$; B is the axion field which in our case 
is zero and $b$ is related to the one-loop $\b$-function coefficients
associated with the $N=2$ subsection \cite{ka}.
Note that for the abelian gauge fields which we consider, the constant $b$ 
is negative. Using the  modular property of the $\eta$-function
$\eta(-\frac{1}{\tau})=(-i\tau)^{1/2}\eta(\tau)$ one can check that
the choice  $G_{11}=fk$ and $G_{22}=FK$ yields the same result for $\D$;
\begin{equation}
\label{delta}
\Delta=-b\ln fk \Bigl|\eta(i\sqrt{FKfk})\Bigr|^4
\Bigl|\eta(i\sqrt{\frac{fk}{FK}})\Bigr|^4.
\end{equation}
Some comments on the properties of the threshold corrections are in order. For
the purely electric $(f=k^{-1}=1)$ extremal $(F^{-1}=K)$ black hole as well
as for the purely magnetic $(F^{-1}=K=1)$ extremal $(f=k^{-1})$ black hole,
the threshold correction is constant. The same is true  if
all the four charges are equal $(F^{-1}=K=f=k^{-1})$.  

For the solutions we consider below the threshold correction tends to a finite
nonzero value as the radius $r$ goes to zero and infinity. For 
intermediate values of $r$, the threshold correction is a slowly varying
function of $r$.  This
provides a motivation for considering constant threshold
corrections to these  black holes. 
  For  small $r$ we are
near the horizon of the tree-level solution and for asymptotically large $r$
we obtain the mass and charges of the black hole.  
  It seems natural to consider also near extremal 
black holes $(\D Q\ll Q)$ where the threshold corrections can be developed 
in a power series in $\frac{\D Q}{Q}$ near the
corresponding extremal values.  

\section{Black hole geometry with genus-one corrections}

The equations of motion obtained from the action (\ref{oneloop}) are:
\begin{eqnarray}
\label{eqone}
0&=& g^{\nu\mu}D_{\nu}\pd\p +\frac{1}{4}e^{-\p+2\g}(F_{(1)})^2
+\frac{1}{4}e^{-\p+2\s}(F_{(2)})^2  \\
&+&\frac{1}{4}e^{-\p-2\g}(F_{(3)})^2
+\frac{1}{4}e^{-\p-2\s}(F_{(4)})^2 \nonumber \\
0&=& 2g^{\nu
\mu}D_{\nu}\pd\s -\frac{1}{2}e^{-\p+2\s}(F_{(2)})^2
+\frac{1}{2}e^{-\p-2\s}(F_{(4)})^2 \\
&-&\frac{1}{4}\frac{\pa\D}{\pa\s}((F_{(2)})^2+(F_{(4)})^2) \nonumber \\
0&=&2g^{\nu\mu}D_{\nu}\pd\g -\frac{1}{2}e^{-\p+2\g}(F_{(1)})^2
+\frac{1}{2}e^{-\p-2\g}(F_{(3)})^2 \\
&-&\frac{1}{4}\frac{\pa\D}{\pa\g} ((F_{(1)})^2+(F_{(3)})^2) \nonumber \\
0&=&\pd\left(\sqrt{-g}(e^{-\p+2\g}+\D)g^{\mu\mu'}g^{\nu\nu'}F_{\mu'\nu'}^{(1)}\right) \\
0&=&\pd\left(\sqrt{-g}(e^{-\p+2\s}+\D)g^{\mu\mu'}g^{\nu\nu'}F_{\mu'\nu'}^{(2)}\right) \\
0&=&\pd\left(\sqrt{-g}(e^{-\p-2\g}+\D)g^{\mu\mu'}g^{\nu\nu'}F_{\mu'\nu'}^{(3)}\right) \\
0&=&\pd\left(\sqrt{-g}(e^{-\p-2\s}+\D)g^{\mu\mu'}g^{\nu\nu'}F_{\mu'\nu'}^{(4)}\right) \\
0&=&R_{\mu\nu}-\frac{1}{2}g_{\mu\nu}R-\frac{1}{2}\pd\p\pa_{\nu}\p
+\frac{1}{4}g_{\mu\nu}g^{\a\b}(\pa_{\a}\p\pa_{\b}\p-2D_{\a}\pa_{\b}\p) \\
&-&\pd\g\pa_{\nu}\g+\frac{1}{2}g_{\mu\nu}g^{\a\b}\pa_{\a}\g\pa_{\b}\g
-\pd\s\pa_{\nu}\s+\frac{1}{2}g_{\mu\nu}g^{\a\b}\pa_{\a}\s\pa_{\b}\s
-\frac{(e^{-\p+2\g}+\D)}{2}F_{\mu\nu}^{(1)^2} \nonumber \\
&-&\frac{(e^{-\p+2\s}+\D)}{2}F_{\mu\nu}^{(2)^2}
-\frac{(e^{-\p-2\g}+\D)}{2}F_{\mu\nu}^{(3)^2}
-\frac{(e^{-\p-2\s}+\D)}{2}F_{\mu\nu}^{(4)^2}. \nonumber
\end{eqnarray}

Note that the string-tree-level chiral null model provides a  solution to
 the low energy effective action (\ref{ea}) which in a special renormalization 
 scheme does not receive
$\a'$ corrections. Solutions of the one-loop-corrected effective action 
(\ref{oneloop}) modify  the harmonic 
condition for the functions $F^{-1}, K, f, k^{-1}$ which is substantial in
the  proof of the
absence of $\a'$ corrections. The $\a'$ corrections become
important near the horizon which for all the black holes we consider here is
 at $r=0$. If the modified functions
 $F^{-1}, K, f, k^{-1}$ satisfy harmonic equation, we can expect that there 
will be no $\a'$
corrections. Below we shall see  that this can be the case for some
black holes, for example, for the extremal dyonic solution.

\subsection{ Magnetic extremal black hole}

In this subsection we consider purely magnetic
$(F^{-1}=K=1)$ extremal $(f=k^{-1})$ black holes.
These are one of the few black holes for which the correspondence principle
\cite{hopo} does not apply. The reason lies in the behavior of the dilaton
near the horizon. This "pathology" of magnetic solutions was already noted in
the flat space analysis of  \cite{cvts1}. As  mentioned in the previous 
section, the threshold correction for this black hole  is independent 
of $r$ and equal to
$\D=-b\ln|\eta(i)|^8$ which we will consider  to be numerically small.  
We shall look for
solutions to the equations (8)-(15) in the form:
\begin{eqnarray}
g_{00}&=&-\tilde{\l_1}=-\frac{1}{f_0}-\D\l_1 \nonumber \\
g_{ij}&=&\delta_{ij}\tilde{\l_2}^{-1}=\delta_{ij}(\frac{1}{f_0}+\D\l_2)^{-1}
\nonumber \\
\p&=&\ln f_0 +\D \p_1 \nonumber \\
\g&=&0 \\
F_{ij}^{(1)}&=&F_{ij}^{(3)}=-\epsilon_{ijk}\pa^k(f_0+\D f_1)
\end{eqnarray}
Here $f_0$ is a solution of harmonic equation 
 $\delta^{ij}\pa_i\pa_j f_0=0$.
 In the first order in $\D$ equations are
\begin{eqnarray}
0&=&\pa^{2}\p_1+\frac{\l_1+\l_2}{2f_0}(\pa f_0)^{2}
-\frac{\p_1}{f_0^2}(\pa f_0)^{2}-\frac{1}{2}\pa f_0\pa(\l_2-\l_1) \\
&+&\frac{2}{f_0^2}\pa f_0\pa f_1 \nonumber \\
0&=&\frac{1}{2}\pa^{2}\p_1+f_0\pa^2\l_2+\frac{9}{4}\pa f_0\pa\l_2
+\frac{1}{4}\pa f_0\pa\l_1-\frac{1}{2f_0}\pa f_0\pa\p_1\\
&+& \frac{\l_1+\l_2}{4f_0}(\pa f_0)^2 \nonumber \\
0&=&\frac{f_0}{2}\pa_i\pa_j(\l_2-\l_1)+\frac{2}{f_0^2}\pa_{(i} f_0\pa_{j)} f_1
-\frac{1}{f_0}\pa_{(i} f_0\pa_{j)} \p_1 +\pa_{(i} f_0\pa_{j)}\l_2 
-\frac{\p_1}{f_0^2}\pa_i f_0 \pa_j f_0\\
&+&\l_1(\frac{1}{f_0}\pa_i f_0 \pa_j f_0-\frac{1}{2}\pa_i\pa_j f_0)
+\l_2(\frac{1}{f_0}\pa_i f_0\pa_j f_0+\frac{1}{2}\pa_i\pa_j f_0)
+\frac{1}{f_0}\pa_i f_0 \pa_j f_0\nonumber \\
0&=&\frac{f_0}{2}\pa^2(\l_2-\l_1)+\frac{1}{2}\pa^{2}\p_1
+\frac{2}{f_0^2}\pa f_0\pa f_1-\frac{1}{2f_0}\pa f_0\pa \p_1
+\frac{3}{4}\pa f_0\pa\l_2 \\
&-&\frac{1}{4}\pa f_0\pa\l_1
-\frac{\p_1}{f_0^2}(\pa f_0)^2+\frac{3(\l_1+\l_2)}{4f_0}(\pa f_0)^2
+\frac{1}{f_0}(\pa f_0)^2,
\end{eqnarray}
here the first equation is  the dilaton equation of motion (8), the second
equation is the 
$(00)$ components  of equation (15), and the  last two equations are the 
 transverse and
longitudinal parts  of the $(ij)$ component of equation (15)
The equations for  gauge fields are satisfied if we assume that all fields
depend on $r$ through $f_0$. It is convenient to rewrite the equations in 
 new variables 
\begin{eqnarray}
\psi&=&\frac{\l_1+\l_2}{2}f_0-\p_1 \nonumber \\
\tilde{\l}&=&\l_2-\l_1\\
\end{eqnarray}
 Integrating the system in these new variables effectively requires solving 
first-order differential equations. Integration constants are defined so
that for large values of $r$ the metric is asymptotic to the Lorentz metric.
We obtain
\begin{eqnarray}
\p&=&\ln f_0+\frac{\D}{2}f_0(\ln f_0-1) \nonumber \\
f&=&f_0-\frac{\D}{2}f_0^2 \nonumber \\
\tilde{\l_1}&=&\frac{1}{f_0}+\frac{3\D}{2}\ln f_0 \nonumber \\
\tilde{\l_2}&=&\frac{1}{f_0}-\frac{\D}{2}\ln f_0.
\end{eqnarray}
Similar logarithmic corrections to the geometry of black holes
appeared in a completely different quantum gravity approach in \cite{so}.
The asymptotic charges are
\begin{eqnarray}
\label{charge}
M&=&\frac{P}{2}(1-\frac{3\D}{2}) \nonumber \\
Q_{m}&=&P(1-\D) \nonumber \\
D&=& P
\end{eqnarray}
where $D$ is the dilaton charge. Although the tree-level solution was BPS saturated,
 the genus-one-corrected solution is no
longer BPS saturated: $M\ne Q_m$. The location of the horizon  is found from the
equation 
\begin{equation}
\frac{r}{P+r}-\frac{3}{2}|\D|\ln(1+\frac{P}{r})=0.
\end{equation}
(note that  $\D<0)$.
From this  transcendental equation we find that for small values of $\D$ 
the horizon is located at $r=3|\D| P/2\ln(2/3|\D|)$ \cite{olver} 
which is close to zero. We see 
that that horizon is pushed outward and the mass is increased as in the case 
of the electric black hole considered in \cite{chan}. It is interesting to
note that
if the black hole we consider were non abelian $b>0$  yielding
positive $\D>0$ (note that $\ln T|\eta(iT)|^4$ is always negative), 
 there would be no horizon. 

\subsection{ Dyonic extremal black hole}

Here we consider the case where all the charges are equal
($F^{-1}=K=f=k^{-1}$), now  all the moduli  as 
well as the dilaton are zero (see definitions (\ref{back})).
Before beginning calculations similar to the those of the previous
subsection, note that the one-loop-corrected effective action is now of the form
\begin{equation}
\label{dyon}
S=\int d^4x \sqrt{-g}\left(R-\frac{1+\D}{2}(F_1^2+F_2^2)\right).
\end{equation}
From this expression it is clear that a solution can be found by redefining 
 electric  and magnetic fields and 
leaving the metric intact. Namely
\begin{eqnarray}
F^{(1)}_{ij}&=&-\epsilon_{ijk}\pa^k \frac{f_0}{\sqrt{1+\D}} \nonumber \\
F^{(2)}_{0i}&=&\pa_i(\sqrt{1+\D}f_0)^{-1} 
\end{eqnarray}
After this redefinition the action (\ref{dyon}) takes the form of the 
tree-level effective action. Decomposing the metric and gauge  fields as
in the previous subsection, we obtain other solutions to (\ref{dyon}),
but only (29) is  asymptotic to the solution
 of the tree-level effective action.
The requirement that both solutions have the same asymptotics is natural 
from the physical point of view, because
small perturbation of the couplings should not affect  the spacetime
 far  from the black hole. The electric and magnetic charges are
 modified in a different way: $Q_{e}=\sqrt{1+\D} Q$ and  $Q_{m}=Q/\sqrt{1+\D}$.
One can see that the BPS condition is 
violated (note, however, that in the first order in $\D$  
$M=Q=\frac{1}{2}(Q(1+\D/2)+Q(1-\D/2))$). 
Although the charges are modified,  the 
entropy is unchanged.
The position of the horizon is again at $r=0$. 
Since the modified potentials are obtained by rescaling  the tree-level
ones, they satisfy harmonic equation and receive no $\a'$ corrections. 

\subsection {Electric extremal black hole}

For the purely electric  $(f=k^{-1}=1)$ extremal $(F^{-1}=K)$ black hole,
using the following ansatz

\begin{eqnarray}
F_{0i}&=&\pa_i K_0^{-1} +\D\pa_i K_1 \nonumber \\
g_{00}&=&-(\frac{1}{K_0}+\D\l_1) \nonumber \\
g_{ij}&=&\delta_{ij}(\frac{1}{K_0}+\D\l_2)^{-1}\nonumber \\
\p&=&-\ln K_0 +\D\p_1,
\end{eqnarray}
where all functions depend only on the $r$, one finds that there is no
solution to the one-loop effective action. The obstruction is
connected with the fact that for this ansatz the equation for the $g^{ij}$
components of the metric gives two equations since there are two
different tensor structures ($\pa_i K \pa_j K$ and $\delta_{ij}(\pa K)^2$).
In the magnetic case, because of the presence of the antisymmetric tensor in the
definition of magnetic field, the system of equations of motion can be
satisfied if all the fields depend on  $r$ through $f_0$.
  In the electric case the gauge field
equation
is not so easily satisfied and yields a system of equations that 
has no solutions for the spherically symmetric configuration.
 More exactly, for the ansatz with
the special part of the metric proportional to $\delta_{ij}$ we have five
independent
equations and four variables. The same is true for a general
spherically-symmetric configuration in the magnetic case without the
special assumption
that  $r$-dependence is through  $f_0$. We conclude that for the electric extremal
black hole the one-loop correction breaks the the spherical symmetry of the
tree-level solution.

\section{Genus-one-corrected magnetic black hole and the correspondence
principle}
 
The reason why the correspondence principle \cite{sus,ho,hopo} 
cannot be applied to  magnetic black hole
is  the behavior of the string coupling constant $g_{st}=e^{\p}$
as one approaches the
horizon. At the tree level, we have

\begin{equation}
g_{st}=f_0=1+\frac{P}{r}
\end{equation}
and for $r\to 0$, $g_{st}\to \infty$. Using the one-loop-corrected dilaton,
we find that now the coupling constant is

\begin{equation}
g_{st}=\exp(\p)=f_0\exp\left(-\frac{|\D|}{2}f_0(\ln f_0-1)\right).
\end{equation}
Now as one approaches the horizon which is close to $r=0$, the coupling
constant decreases. This is the behavior required for applicability of
the correspondence principle. In this section  we will show that using the
correspondence principle one can find a microscopic description of the
entropy of 4D magnetic black hole. The classical (macroscopic) entropy of the
magnetic black hole, defined  as one fourth of the area of the horizon is
\begin{equation}
S=\frac{A}{4}=\pi(\l_2^{-1}r^2)|_{hor}=\frac{9}{8}\pi|\D|P^2\ln(2/3|\D|)
\end{equation}

String tree-level
black hole solution  which is represented by the "chiral null model" can be
embedded both in NS-NS \cite{ts2} and R-R sectors \cite{rr} of string
theory. Let us consider the R-R embedding in type IIB theory. We use the
one-loop-corrected magnetic black hole solution as background in bosonic
part of the 2D world-sheet action of IIB superstring theory. Conformal
invariance of this theory is preserved if we consider both tree-level and
one-string-loop contributions to the effective action. The bosonic part of the
latter is precisely the action (\ref{oneloop}) which equations of motion were 
solved
to obtain the deformed, i.e. loop-corrected backgrounds.
 
The  first step to understand the microscopic origin of the entropy of magnetic 
black hole is to
represent the 4D magnetic black hole as a solitonic brane.
For this purpose
we  modify the chiral null model in 5D so that under compactification
to 4D it reproduces the one-loop-corrected magnetic black hole. The 4D
background is

\begin{eqnarray}
ds^2_{4D}&=&-\tilde{\l_1}dt^2+\tilde{\l_2}^{-1}dx^idx^i \nonumber \\
\tilde{\l_1}&=&\frac{1}{f_0}-\frac{3|\D|}{2}\ln f_0 \nonumber \\
\tilde{\l_2}&=&\frac{1}{f_0}+\frac{|\D|}{2}\ln f_0 \nonumber \\
\p_{4D}&=&\ln f_0-\frac{|\D|}{2}f_0(\ln f_0-1) \nonumber \\
F_{ij}&=&-\epsilon_{ijk}\pa^k(f_0+\frac{|\D|}{2}f_0^2)
\end{eqnarray}
The 5D background which  compactified to 4D provides the
one-loop-corrected magnetic black hole is
\begin{eqnarray}
\label{5d}
ds^2&=&g_{00}dt^2+dy^2 +2a_idydx^i +(a_ia_j
+\tilde{f}^2\delta_{ij})dx^idx^j \nonumber \\
B_{yi}&=&a_i\nonumber \\
\p&=&\p_{4D} \nonumber \\
a_idx^i&=&P(1+|\D|+\frac{|\D|P}{r})(1-\cos\theta)d\varphi \nonumber \\
g_{00}&=&-(1-\frac{3|\D|}{2}f_0\ln f_0)
\exp(-\frac{|\D|}{2}f_0(\ln f_0-1)) \nonumber \\
\tilde{f}^2&=&\frac{2f_0^2}{2+|\D|f_0\ln f_0}
\exp(-\frac{|\D|}{2}f_0(\ln f_0-1)). \nonumber \\
\end{eqnarray}
Here we have considered the extremal case so that the dilaton in both
dimensions is the same.  
 The obtained configuration is
a solitonic brane with the RR gauge field $B$. The mass, charge and entropy
of this solution are (we omit the numerical factors)
\begin{eqnarray}
M&\sim & \frac{|\D|P}{g^2\a'^4}  V \nonumber \\
Q&\sim & \frac{(1+|\D|)P}{g\a'^{1/2}} \nonumber \\
S&\sim & \frac{|\D|\ln(2/3|\D|)P^2}{g^2\a'^4} V
\label{E1}
\end{eqnarray}
where $V$ is the volume of the six-dimensional torus formed by $y$ and
other five compact coordinates that  in (\ref{5d}) we did not write
explicitly.  This is
still a  strong-coupling description; to find the weak-coupling description
we apply the correspondence  principle. At weak coupling the solitonic
brane,
now with the required behavior of the dilaton, transforms into a state of
D-branes. Following \cite{hopo} we assume that at the transition
point the square of the radius of the horizon is of order $\a'$ and 
the masses of the solitonic brane and D-brane configuration  are equal.
\begin{equation}
r_{hor}^2\sim \a' \qquad M_{D}\sim M
\end{equation}
Under these assumptions we consider a system of $Q$ D branes with a
small number of excited long strings. In this case the entropy is
\cite{hopo}
\begin{eqnarray}
S_{D}&\sim &\a'^{1/2} \delta E \nonumber \\
\delta E &\sim & E-\frac{QV}{g\a'^{7/2}}.
\end{eqnarray} 
Here $E$ is the  energy of the D-brane system which at the transition point 
is equal to the
mass of the solitonic brane, the charges of the solitonic brane and the
D-brane system  are also equal. Substituting the
expression of the energy, we find the entropy of the excited D-brane system:
\begin{equation}
S_{D}\sim \frac{QV}{(1+|\D|)g\a'^3}\sim
\frac{V}{g^2(|\D|\ln(2/3|\D|))^7P^6}
\end{equation}
This is exactly the same expression for the entropy as that of the solitonic 
brane (\ref{E1}) since
\begin{equation}
S\sim  \frac{|\D|\ln(2/3|\D|)P^2}{g^2\a'^4} V
\sim \frac{V}{g^2(|\D|\ln(2/3|\D|))^7P^6}
\end{equation}

Exciting the D-brane system by a large number of massless open strings on
the D-branes \cite{hopo} gives an entropy that does not match that of the
solitonic brane. Probably, this is connected to the fact that although the 
solitonic brane we discussed
here behaves in many respects as a black p-brane, it is not a p-brane, but
 a kind of a hybrid \cite{bh}.

Usually it is expected that unless $\a'$-corrections are absent because of
special properties of solution (such as $N=4$ supersymmetry of the
world-sheet action \cite{cvts1}), these corrections become important near the
singularity and modify the leading term of solution. Qualitative examples of
such modifications for the fundamental string were considered in \cite{ts3}.
The effect of the $\a'$-corrections was argued to smear the singularity and
to move the horizon from $r=0$ to a distance of order $(\a')^{1/2}$, thus
providing a realization of the "stretched horizon" \cite{sugl}. A remarkable
property of the string-loop-corrected magnetic black hole is that now
string-loop correction modifies solution in the same way: it moves the
horizon from $r=0$  and drastically changes the behavior of dilaton which
becomes regular at $r=0$\footnote{For magnetic black hole
string-loop-corrected solution the function $\tilde{f}$ no longer satisfies
the harmonic equation. This results in appearance of $\a'$-corrections which
modify the solution further. We ignore these modifications supposing that they
do not modify the solution qualitatively.}.The latter property was 
crucial for applicability of the correspondence principle.

\section{Conclusions}

We investigated charged extremal black-hole solutions of equations of motion
of the one-string-loop-corrected effective action which provide backgrounds
for the "chiral null model" \cite{bh}.

 In  the purely magnetic case we found an explicit analytical solution 
for the extremal black hole. It appears that geometry is
modified; namely, background  receives logarithmic corrections which are in agreement
with corrections found in a  different approach to quantum gravity \cite{so}. 
The threshold corrections change the values of the mass and magnetic
charge of the black hole  and  move the horizon from $r=0$ to a distance
 proportional to the threshold correction and the charge. 

For the extremal dyonic black hole, 
we found 
solution which has the form of the tree-level solution with the rescaled
electric and magnetic fields. It follows that since the modified
backgrounds satisfy harmonic equation they receive no $\alpha'$ corrections
\cite{cvts1,bh}. The
entropy of the dyonic black hole remain the same as at the tree level,
although the charges are modified. 

For the purely electric black
hole, we found that equations of motion with  threshold corrections  have no
spherically symmetric solution. The same is true for generic
spherically-symmetric magnetic solution, but in the latter case 
because of the special form of magnetic field
 it was possible to construct a particular solution.

Of special interest is application of the correspondence principle to the
loop-corrected solution for the extremal magnetic black hole.
In this case, loop corrections modify the tree-level solution so that the dilaton
which, at the tree-level, is singular at the horizon,
becomes a smooth function which vanishes at $r=0$. This means that the
string coupling is weak near the horizon and opens the
possibility  to apply the correspondence principle to semi-quantitative
microscopic calculation of the entropy of the magnetic black hole.  
It was shown that in the picture in which the weak-coupling D-brane system
is described by a small number of long strings the Bekenstein-Hawking
entropy of magnetic black hole is reproduced by the D-brane system. 

\begin{center}
\begin{Large}
Acknowledgments
\end{Large}
\end{center}

This investigation was partially supported by RFFR grant 96-02-16413.
L.A.P.Z. is very grateful to N. Girard.

\pagebreak

\end{document}